\documentstyle[11pt,amssymb]{article}

\makeatletter
\@addtoreset{equation}{section}
\makeatother

\def\dalemb#1#2{{\vbox{\hrule height .#2pt
        \hbox{\vrule width.#2pt height#1pt \kern#1pt
                \vrule width.#2pt}
        \hrule height.#2pt}}}

\def\hA{\hat{\cal A}}
\def\cF{{\cal F}}
\def\cA{{\cal A}}

\def\p{{\sst{(p)}}}
\def\0{{\sst{(0)}}}
\def\1{{\sst{(1)}}}
\def\2{{\sst{(2)}}}
\def\3{{\sst{(3)}}}
\def\4{{\sst{(4)}}}
\def\5{{\sst{(5)}}}
\def\6{{\sst{(6)}}}
\def\7{{\sst{(7)}}}
\def\8{{\sst{(8)}}}

\def\tD{\widetilde D}
\def\bA{\bar{\cal A}}
\def\bF{\bar{\cal F}}

\def\Z{\rlap{\sf Z}\mkern3mu{\sf Z}}
\def\R{\rlap{\rm I}\mkern3mu{\rm R}}

\def\td{\tilde}
\def\wtd{\widetilde}

\let\a=\alpha \let\b=\beta \let\g=\gamma

\def\nn{\nonumber} \def\bd{\begin{document}} \def\ed{\end{document}}
\def\ds{\documentstyle} \let\fr=\frac \let\bl=\bigl \let\br=\bigr
\let\Br=\Bigr \let\Bl=\Bigl 
\let\bm=\bibitem
\let\na=\nabla
\let\pa=\partial \let\ov=\overline 
\newcommand{\be}{\begin{equation}} 
\newcommand{\ee}{\end{equation}} 
\def\ba{\begin{array}}
\def\ea{\end{array}}
\def\ft#1#2{{\textstyle{{\scriptstyle #1}\over {\scriptstyle #2}}}}
\def\fft#1#2{{#1 \over #2}}
\def\del{\partial}
\def\sst#1{{\scriptscriptstyle #1}}
\def\oneone{\rlap 1\mkern4mu{\rm l}}
\def\ie{{\it i.e.\ }}
\def\via{{\it via}}
\def\semi{{\ltimes}}
\def\v{{\cal V}}
\def\str{{\rm str}}
\def\Dm{{{D_{\sst{max}}}}}

\newcommand{\ho}[1]{$\, ^{#1}$}
\newcommand{\hoch}[1]{$\, ^{#1}$}
\newcommand{\bea}{\begin{eqnarray}} 
\newcommand{\eea}{\end{eqnarray}} 
\newcommand{\ra}{\rightarrow}
\newcommand{\lra}{\longrightarrow}
\newcommand{\Lra}{\Leftrightarrow}
\newcommand{\ap}{\alpha^\prime}
\newcommand{\bp}{\tilde \beta^\prime}
\newcommand{\tr}{{\rm tr} }
\newcommand{\Tr}{{\rm Tr} } 
\newcommand{\NP}{Nucl. Phys. }
\newcommand{\tamphys}{\it Center for Theoretical Physics,
Texas A\&M University, College Station, Texas 77843\\
and SISSA, Via Beirut No. 2-4, 34013 Trieste, Italy\hoch{3}}
\newcommand{\ens}{\it Laboratoire de Physique Th\'eorique de l'\'Ecole
Normale Sup\'erieure\hoch{3,4}\\
24 Rue Lhomond - 75231 Paris CEDEX 05}
\newcommand{\upenn} {\it Dept. of Phys. \& Astro., University of
Pennsylvania, Philadelphia, PA 19104}

\newcommand{\auth}{E. Cremmer\hoch{\dagger}, B. Julia\hoch{\dagger}, 
H. L\"u\hoch{\ddagger1} and C.N. Pope\hoch{\star2}}

\begin{document}
\begin{flushright}
\hfill{LPTENS-99/21}\ \  
\hfill{UPR/858-T} \ \
\hfill{CTP TAMU-35/99}\ \  
\hfill{SISSARef-111/99/EP}\\ 
\hfill{\bf hep-th/9909099}\\
\hfill{September 1999}\\
\end{flushright}

%\vspace{15pt}

\begin{center}
{ \large {\bf Higher-dimensional Origin of $D=3$ Coset Symmetries}}

\vspace{10pt}
\auth

\vspace{10pt}

{\hoch{\dagger}\ens}

\vspace{10pt}
{\hoch{\ddagger}\upenn}

\vspace{10pt}
{\hoch{\star}\tamphys}

\vspace{10pt}

\underline{ABSTRACT}
\end{center}

    It is well known that the toroidal dimensional reduction of
supergravities gives rise in three dimensions to theories whose
bosonic sectors are described purely in terms of scalar degrees of
freedom, which parameterise sigma-model coset spaces.  For example, the
reduction of eleven-dimensional supergravity gives rise to an
$E_8/SO(16)$ coset Lagrangian. In this paper, we dispense with the
restrictions of supersymmetry, and study all the three-dimensional
scalar sigma models $G/H$ where $G$ is a maximally-non-compact simple
group, with $H$ its maximal compact subgroup, and find the highest
dimensions from which they can be obtained by Kaluza-Klein reduction.
A magic triangle emerges with a duality between rank and dimension.
Interesting also are the cases of Hermitean symmetric spaces and
quaternionic spaces.

{\vfill\leftline{}\vfill
%\vskip 5pt
\footnoterule
{\footnotesize  \hoch{1} Research supported in part by DOE 
Grant DE-FG02-95ER40893 \vskip  -12pt} \vskip 14pt
{\footnotesize  \hoch{2} Research supported in part by DOE
Grand DE-FG03-95ER40917 \vskip  -12pt} \vskip 14pt
{\footnotesize \hoch{3} Research supported in part by EC under TMR
contract ERBFMRX-CT96-0045 \vskip -12pt} \vskip 14pt
{\footnotesize
        \hoch{4} UMR 8549 du Centre National de la Recherche
Scientifique et de l'\'Ecole Normale Sup\'erieure \vskip -12pt}
                       \vskip 10pt
%{\footnotesize \hoch{\phantom{3}} et \`a l'Universit\'e de Paris-Sud 
%\vskip -12pt}}

%\baselineskip=24pt
\pagebreak
\setcounter{page}{1}

\tableofcontents
\addtocontents{toc}{\protect\setcounter{tocdepth}{2}}
\newpage

\section{Introduction}

    It is well known that the dimensional reduction of eleven-dimensional
supergravity \cite{cjs} to $D=3$ gives rise to a scalar coset theory
with an $E_{8(+8)}$ global symmetry.  Conversely, we may say that the
``oxidation endpoint'' of the three-dimensional $E_{8(+8)}$ scalar
coset is the bosonic sector of $D=11$ supergravity.  In \cite{bgm}, a
classification of three-dimensional scalar cosets associated with the
various groups that arise from the dimensional reduction of $D=4$
gravity coupled to matter systems was given. In many cases, these
four-dimensional theories can themselves be obtained from the
dimensional reduction of yet higher dimensional theories of matter
coupled to gravity.  In most of this paper, we shall study the ``oxidation
endpoints'' (also called group disintegration endpoints) 
of the three-dimensional scalar cosets given in
\cite{bgm}.  By the ``oxidation endpoint'' of a three-dimensional
scalar coset, we mean the (bosonic) theory in the highest possible dimension
$\Dm$ whose toroidal dimensional reduction gives back precisely the
three-dimensional scalar coset.  We shall concentrate on the examples where
the numerator group of the three-dimensional scalar coset is maximally
non-compact. It turns out that these cases are simpler and exhibit more 
regularity than the cases of non-maximally non-compact symmetry groups first 
studied systematically for four dimensional supergravities in \cite{b81}. 

    A particular example, namely the $SL(n, \R)/O(n)$ scalar coset,
was studied in \cite{cjlp2}.  It was shown that it can be obtained
from the dimensional reduction of pure gravity in $D=2+n$ dimensions.
This in fact provides a useful clue in more general cases as to what
is the oxidation endpoint of a given three-dimensional scalar
coset. Namely, if the global symmetry group of a three-dimensional
scalar coset contains a subgroup $SL(n, \R)$, then one might expect
that the oxidation endpoint of the coset would be in $\Dm=n+2$
dimensions.  We find that this rule works for all the maximally
non-compact simple or semi-simple groups, with the exception of $C_k$.
The reason for this can be clarified by recalling that the reduction
to two dimensions leads to a set of equations invariant under the
Kac-Moody affine symmetry.  This symmetry extends the three-dimensional finite
dimensional symmetry group in such a way that the affine Dynkin
diagram reveals the subgroup $SL(n, \R)$ by direct inspection (for
instance the (Freudenthal-) subgroup $SL(9, \R)$ of $E_8$).  One
can then check that $C_k$ cannot be oxidised beyond 4 dimensions. The point
is that the affine vertex of the extended diagram should be the
endpoint of this $SL(n, \R)$ (see, for instance, \cite{b81}). Let us
also point out that the toroidal dimensional reduction of any
$D$-dimensional generally-covariant theory to $D=3$ will automatically
admit a $GL(D-3,\R)$ rigid internal symmetry.  Part of the magic here
is that $SL(D-2,\R)$ invariance holds, as one can check in a case by
case analysis.

    Having found the oxidation endpoints for the various
three-dimensional scalar cosets, one can then look also at the global
symmetry groups that result when these higher-dimensional theories are
reduced to various intermediate dimensions $ 3<D <\Dm$.  Each
oxidation endpoint theory gives rise to a particular sequence of such
intermediate theories, which we may label by the associated global
symmetry group in $D=3$.  The most well-known example is thus the
``$E_8$ sequence,'' which comes from the dimensional reduction of
$\Dm=11$ supergravity, giving $E_{\sst{11-D}}$ in $D$ dimensions.
Note that, unlike the situation discussed in \cite{bgm,hj,cllpst}, we
shall principally be concerned with the case where the dimensional
reduction is on a purely spacelike torus.  The only difference, from
the group-theoretic point of view, is that the denominator groups $H$
for the scalar cosets $G/H$ will be compact rather than non-compact.

\section{Simply-laced scalar coset sequences \label{sec:simple}}

\subsection{$A_n$ sequences}

    The maximally non-compact form of $A_n$ is $SL(n+1,\R)$.  As
mentioned in the Introduction, the associated three-dimensional scalar
cosets $SL(n+1,\R)/O(n+1)$ can be obtained by the dimensional
reduction of pure gravity in $\Dm=n+3$ dimensions \cite{cjlp2}.

   If we perform the dimensional reduction to $D=3$  of the pure
Einstein-Hilbert Lagrangian in $\Dm=n+3$, we obtain
%%%%%
\be
{\cal L}_3 =  R\, {*\oneone}  
-\ft12 \, {*d\vec\phi}\wedge d\vec\phi -\ft12  \sum_i e^{\vec
b_i\cdot\vec\phi}\, {*{\cal F}_\2^i}\wedge \cF_\2^i -\ft12 
\sum_{i<j} e^{\vec b_{ij}\cdot\vec\phi}\, {*{\cal F}_{\1 j}^i} \wedge
\cF^i_{\1 j}\ .
\label{einstred}
\ee
%%%%%
After dualising the potentials $\cA_\1^i$ to give further axions
$\chi_i$, one obtains the scalar Lagrangian
%%%%
\be
{\cal L}_3 = R\, {*\oneone}  
-\ft12 \, {*d\vec\phi}\wedge d\vec\phi
-\ft12 \sum_i e^{-\vec
b_i\cdot\vec\phi}\, {*G_{\1 i}}\wedge G_{\1 i} -\ft12 
\sum_{i<j} e^{\vec b_{ij}\cdot\vec\phi}\, {*{\cal F}_{\1 j}^i} \wedge
\cF^i_{\1 j}\ ,
\label{einstred2}
\ee
%%%%%
where ${*{\cal F}_\2^i}\, e^{\vec b_i\cdot\vec\phi} = G_{\1 i}\equiv
\gamma^j{}_i\, d \chi_j$.  It is now evident that we can extend the
range of the $i$ index to $a=(0,i)$ (with $0<i$ here), and define
axions $\bA^a_{\0 b}$ for all $a<b$:
%%%%%%%%
\be
\bA^0_{\0 i} = \chi_i\ ,\qquad \bA^i{}_{\0 j} = {\cal A}^i_{\0 j}\ .
\ee
%%%%%%%%
(The bar over the potential indicates the extended set.)  Defining
also $\bar\g^a{}_b$ as in Appendix A, but for the
extended set of axionic potentials $\bA^a{}_{\0 b}$, by $\bar\g^0{}_i=
-\chi_i$ and $\bar\g^i{}_j =\g^i{}_j$, we see that (\ref{einstred2})
assumes the form
%%%%%%%%
\be
{\cal L} = R\, {*\oneone}
-\ft12 e\, {*d\vec\phi}\wedge d\vec\phi
 -\ft12 
\sum_{a<b} e^{\vec b_{ab}\cdot\vec\phi}\, {*\bF_{\1 b}^a}
\wedge \bF_{\1 b}^a\ ,
\label{einstred3}
\ee
%%%%%%%
Thus the usual proof of the existence of the $SL(n,\R)$ symmetry now
establishes that we have an $SL(n+1,\R)$ symmetry in this
three-dimensional case (see, for example, \cite{cjlp2}).

    The dilaton vectors $\vec b_{ij}$ and $-\vec b_i$ form the
positive roots of $SL(n+1,\R)$, and it is easily seen that $-\vec b_1$
and $\vec b_{i,i+1}$ (with $1\le i\le n-1$) are the simple roots.
Thus we have the Dynkin diagram \cite{cjlp2}

\centerline{
\begin{tabular}{ccccccccccc}\\
 $-\vec b_{1}$& &$\vec b_{12}$&&$\vec b_{23}$ & && &$\vec b_{n-2,n-1}$
& &$\vec b_{n-1,n}$ \\
 $\scriptstyle{\bigcirc}$ &---& $\scriptstyle{\bigcirc}$
  &---& $\scriptstyle{\bigcirc}$&
---& $\cdots\cdots$&---&$\scriptstyle{\bigcirc}$&---&
   $\scriptstyle{\bigcirc}$ \\
\end{tabular}}
\bigskip\bigskip

\begin{itemize}

\item[] {\bf Diagram 1}: In $D=3$, $-\vec b_1$ and
$\vec b_{i,i+1}$ generate the $SL(n+1,\R)$ Dynkin diagram. Vertices
appear one by one from the right, starting 
in dimension $1+n$, except for the left-most one, which 
is a disconnected $\R$ from dimensions $2+n$ down to
$3$, where it connects as an $SL(2, \R)$ subgroup. 
\end{itemize}

\bigskip\bigskip
 
   If the pure gravity theory in $\Dm=n+3$ dimensions is reduced to
intermediate dimensions 
$D= \Dm-m >3$ on an $m$-torus, then the global symmetry group will be
$GL(m,\R)$.

\subsection{$D_n$ sequences}

    The maximally non-compact form of $D_n$ is $O(n,n)$.  These groups
are familiar as the T-duality symmetry groups of string theory,
dimensionally reduced on a torus.  For example, if one reduces the
NS-NS sector of ten-dimensional type II string to $D\ge 5$ on
$T^{10-\sst D}$, one obtains an $O(10-D,10-D)\times \R$ global
symmetry.  There are symmetry enhancements in $D=4$ and $D=3$, leading
to $O(6,6)\times SL(2,\R)$, and $O(8,8)$ respectively
\cite{b85,sen1,sen2}. More generally, we can consider the Lagrangian
%%%%%
\be
{\cal L}_{\sst D} =R\, {*\oneone} - \ft12 {*d\varphi}\wedge d\varphi -
\ft12 e^{a\varphi}\, {*F_\3}\wedge F_\3 \label{dnlag}
\ee
%%%%%
in an arbitrary dimension $\Dm$, where the constant $a$ is given by
$a^2=8/(\Dm-2)$.  Upon toroidal reduction to
$D=3$, we obtain
%%%%%
\bea
{\cal L}_3 &=& R\, {*\oneone} - \ft12 {*d\vec\phi}\wedge d\vec\phi -
\ft12 \sum_i e^{\vec b_i\cdot\vec\phi}\, {*\cF_\2^i}\wedge \cF_\2^i 
-\ft12 \sum_{i<j} e^{\vec b_{ij}\cdot\vec\phi}\, {*\cF^i_{\1 j}} \wedge  
\cF^i_{\1 j}\nn\\
&&-\ft12 \sum_i e^{\vec a_i\cdot\vec\phi}\, {*F_{\2 i}}\wedge F_{\2 i}
-\ft12 \sum_{i<j} e^{\vec a_{ij}\cdot\vec\phi}\, {*F_{\1 ij}}\wedge 
F_{\1 ij}\ .\label{dnlag1}
\eea
%%%%%
Note that here $\vec\phi$ denotes the set of dilatons
$(\phi_1,\phi_2,\ldots, \phi_{{\sst \Dm}-3})$ introduced in Appendix A,
augmented by $\varphi$ as a zero'th component; $\vec\phi=(\varphi,
\phi_1,\phi_2,\ldots, \phi_{{\sst \Dm}-3})$.  Similarly, the various
dilaton vectors are those defined in Appendix A, augmented by a first
component that is equal to the constant $a$ in the case of $\vec a_i$
and $\vec a_{ij}$, and is equal to zero in the case of $\vec b_i$ and
$\vec b_{ij}$. 

    After dualising the 1-form potentials $\cA^i_\1$ and $A_{\1 i}$ to
axions $\chi_i$ and $\psi^i$ respectively, the
three-dimensional Lagrangian (\ref{dnlag1}) can written as the purely
scalar Lagrangian 
%%%%%
\bea
{\cal L}_3 &=& R\, {*\oneone} - \ft12 {*d\vec\phi}\wedge d\vec\phi -
\ft12 \sum_i e^{-\vec b_i\cdot\vec\phi}\, {*G_{\1 i}}\wedge G_{\1 i} 
-\ft12 \sum_{i<j} e^{\vec b_{ij}\cdot\vec\phi}\, {*\cF^i_{\1 j}} \wedge  
\cF^i_{\1 j}\nn\\
&&-\ft12 \sum_i e^{-\vec a_i\cdot\vec\phi}\, {*G_\1^i}\wedge G_\1^i
-\ft12 \sum_{i<j} e^{\vec a_{ij}\cdot\vec\phi}\, {*F_{\1 ij}}\wedge 
F_{\1 ij}\ ,\label{dnlag2}
\eea
%%%%%
where the dualised field strengths are given by
%%%%%
\bea
G_{\1 i} &=& \gamma^j{}_i\, (d\chi_j - A_{\0 kj}\, d\psi^k)\ ,\nn\\
G^i_\1 &=& \td\gamma^i{}_j\, d\psi^j\ .
\eea
%%%%%
(See Appendix A for details of how the dualisations work.)

     It is now straightforward to see that if we take $\Dm=n+2$, then
the three-dimensional Lagrangian (\ref{dnlag2}) has a $D_n$ global
symmetry, where $(\vec b_{ij}, -\vec b_i, \vec a_{ij}, -\vec a_i)$ are
its positive roots, with the simple roots being $\vec a_{12}$, $\vec
b_{i,i+1}$ ($i\le n-1$) and $-\vec a_n$.  Thus we have the Dynkin
diagram

\bigskip\bigskip
 
\centerline{
\begin{tabular}{ccccccccccc}\\
$\vec b_{12}$& &$\vec b_{23}$& &$\vec b_{34}$&
& & & $\vec b_{n-1, n}$ & & $-\vec a_{n}$\\
 o&---&o&---&o&---&$\cdots\cdots$ & --- &o & --- &o\\
   &   &$|$&   & &   & &&&&\\
   &   &o&   & &   & &&&& \\
   &   &$\vec a_{12}$& & &&&&&   &  \\
\end{tabular}}
\bigskip
 
\begin{itemize}
\item[]{\bf Diagram 2}: The dilaton vectors $\vec a_{12}$, $\vec
b_{i,i+1}$ and $-\vec a_{n}$ generate the $D_n$ Dynkin diagram. The
vertices appear one by one, starting from the extremities of the horizontal
line which are both $\R$ factors in dimension $n+1$.  The $SL(2,\R)$ on
the left appears in dimension $n$, the short leg being still
$\R$, and the inner roots appear from the left.  The diagram
becomes irreducible in dimension $3$. 
\end{itemize}

\bigskip\bigskip

    In other words, we have shown that a three-dimensional scalar
coset Lagrangian with $G/H= O(n,n)/(O(n)\times O(n))$ can be
dimensionally oxidised to the endpoint dimension $\Dm=n+2$, where it
is described by the Lagrangian (\ref{dnlag}).

    The same Dynkin diagram can be used in order to determine the
global symmetry groups of the intermediate steps of oxidation of the
three-dimensional Lagrangian (\ref{dnlag2}), by removing from the
right those vertices that are associated with fields that exist only
in dimensions lower than the dimension under consideration.  The
vertex $-\vec a_n$ itself must be handled carefully, since it is
associated with a dualised field.  In four dimensions it should be
replaced by $-\vec a$, whilst it corresponds in dimensions greater
than four to the top dimension (dilaton) scalar field.  Consequently,
the global symmetry group of the theory obtained by oxidising the
$D_n$-symmetric three-dimensional scalar Lagrangian to $D\le \Dm=n+2$
dimensions is $D_m\times \R$, where $m= \Dm- D$, if $D\ge 5$, while it
is $D_{n-2}\times SL(2,\R)$ if $D=4$.  (The additional $\R$ factor in
$D\ge5$ is descended from the $\R$ symmetry of the $(n+2)$-dimensional
Lagrangian (\ref{dnlag}); similarly, the Kalb-Ramond field yields a
full $SL(2,\R)$ subgroup connected to the left part of the diagram
from dimension $n-1$ downwards.)  The connection of the KR degrees of
freedom to the second node on the left is due to the fact that it is a
second-rank antisymmetric tensor.  We shall see shortly that a
third-rank RR form leads to the $E_n$ series for a similar reason.

\subsection{$E_n$ sequences}

\subsubsection{$E_8$ sequence}

    This is the well-known sequence of $D$-dimensional scalar coset
theories obtained by the dimensional reduction of eleven-dimensional
supergravity on the $(D-11)$-torus \cite{cjgroup}, whose bosonic
sector is given by
%%%%%
\be
{\cal L}_{11} = R\, {*\oneone} - \ft12 {*F_\4}\wedge F_\4 - \ft16
dA_\3\wedge dA_\3 \wedge A_\3\ .
\ee
%%%%%
After
dualising all higher-degree fields where necessary, the scalar sector
of the $D$-dimensional theory has $E_{11-D}$, in its maximally
non-compact form, as its global symmetry group.  A discussion of these
symmetries, in the language we are using in this paper, may be found
in \cite{cjlp1}.  The scalar fields in $D$ dimensions are the dilatons
$\vec\phi$, the axions $\cA^i_{\0 j}$ coming from the metric and
$A_{\0 ijk}$ coming from the eleven-dimensional 3-form $A_\3$,
together with further axions in $D\le 5$ coming from the dualisation
of certain higher-degree gauge fields.  The simple-root vectors are
$\vec b_{i,i+1}$ and $\vec a_{123}$ (associated with the dilaton
vectors $\vec b_{ij}$ and $\vec a_{ijk}$ for $\cA^i{}_{\0 j}$ and
$A_{\0ijk}$ respectively).  These give the Dynkin diagrams
depicted in Diagram 3:

\bigskip\bigskip

\centerline{
\begin{tabular}{ccccccccccccc}\\
 $\vec b_{12}$& &$\vec b_{23}$& &$\vec b_{34}$& &$\vec b_{45}$
& &$\vec b_{56}$& &$\vec b_{67}$& &$\vec b_{78}$ \\
 o&---&o&---&o&---&o&---&o&---&o&---&o\\
 &   & &   &$|$&   & &   & &   & &   & \\
 &   & &   &o&   & &   & &   & &   & \\
 &   & &   &$\vec a_{123}$& & &   & &   & &   & \\
\end{tabular}}
\bigskip\bigskip

\begin{itemize}
\item[]{\bf Diagram 3}: The dilaton vectors $\vec b_{i,i+1}$ and 
$\vec a_{123}$ generate the $E_8$ Dynkin diagram. The
short leg starts as  $\R$ in $D=10$, becomes $SL(2,\R)$ in
$D=8$, and connects below that.
\end{itemize}
\bigskip\bigskip

   In a given dimension $D$, only those vertices with index values
$i\le 11-D$ survive, giving rise to the Dynkin diagram for
$E_{11-D}$.  (We adopt the standard convention where $E_5\sim D_5$,
$E_4\sim A_4$, $E_3\sim A_2\times A_1$, $E_2\sim GL(2,\R)$ and
$E_1\sim \R$.)

\subsubsection{$E_7$ sequence of scalar cosets}

To obtain this sequence, we consider a consistent (albeit
non-supersymmetric) truncation of $D=9$ maximal supergravity, to the
theory whose bosonic sector comprises just the metric, a dilaton, a
vector potential and a 3-form potential.\footnote{By a ``consistent
truncation,'' we mean that all the equations of motion, including
those of the fields that are set to zero, are satisfied.}  The Lagrangian is
%%%%%
\be
{\cal L}_9 = R\, {*\oneone} -\ft12{*d\varphi}\wedge d\varphi - \ft12
e^{\fft{2}{\sqrt7}\varphi}\, {*F_\4}\wedge F_\4 -\ft12
e^{-\fft4{\sqrt7}\varphi}\, {*F_\2}\wedge F_\2 -\ft12 dA_\3 \wedge dA_\3
\wedge A_\1\ .\label{d9trunc}
\ee
%%%%%
To be precise, the potential $A_\1$, with field strength $F_\2 =
dA_\1$, is the potential $A_{\1 12}$ coming from $A_\3$ after two
steps of toroidal reduction, in the
notation of \cite{lpsol,cjlp1}.  The dilaton $\varphi$ is the linear
combination of the two dilatons of $D=9$ maximal supergravity that is
parallel to $\vec a$ (and hence also parallel to $\vec a_{12}$) in the
notation of \cite{lpsol,cjlp1}.  (In other words, $\varphi = \ft12
\sqrt7 \,\vec a\cdot\vec\phi= -\ft14 \sqrt7\, \vec a_{12}\cdot \vec\phi$.)
However, having noted the $D=11$ origin of the truncated theory, we
shall now use the more convenient notation where we simply denote the
vector potential by $A_\1$.  Further dimensional reductions will use a
notation in the spirit of \cite{lpsol,cjlp1}, but where the first
reduction step, with index $i=1$, is from $D=9$ to $D=8$, and so on.

     Note that in obtaining the Lagrangian (\ref{d9trunc}) the largest
possible number of fields has been consistently truncated out, while
still retaining the 3-form potential $A_\3$. In particular, the $FFA$
term in (\ref{d9trunc}) makes it impossible to make a further
truncation of the vector potential $A_\1$ while keeping $A_\3$.
Another way of understanding the consistency of the truncation is by
looking at maximal $D=9$ supergravity from the point of view of type
IIB supergravity.  The truncation to (\ref{d9trunc}) is then the one
that results from keeping just the metric and the self-dual 5-form
in $D=10$.  In fact, this ten-dimensional theory can also be
viewed as a natural oxidation endpoint for the $E_7$ sequence of theories
that we are considering in this subsection.  The truncation can easily
be seen to be consistent, because all the singlet fields of the
$SL(2,R)$ global symmetry in $D=9$ (or $D=10$ type IIB) are retained.
Note that $SL(8,\R)$ is indeed a subgroup of $E_7$, and so the
$SL(D-2,\R)$ rule mentioned in the Introduction is satisfied.

     Performing the dimensional reduction, using a notation analogous
to that in \cite{lpsol,cjlp1}, we obtain the $D$-dimensional
Lagrangian
%%%%%
\bea
{\cal L}_{\sst D} &=& R\, {*\oneone} - \ft12 {*d\vec\phi}\wedge
d\vec\phi -e^{\vec a\cdot\vec\phi}\, {*F_\4}\wedge F_\4 -\ft12 
\sum_i e^{\vec a_i\cdot\vec\phi}\, {*F_{\3 i}}\wedge F_{\3 i} 
\nn\\
&&-\ft12 \sum_{i<j} 
        e^{\vec a_{ij}\cdot\vec\phi}\, {*F_{\2 ij}}\wedge F_{\2 ij}
-\ft12 \sum_{i<j<k} 
        e^{\vec a_{ijk}\cdot\vec\phi}\, {*F_{\1 ijk}}\wedge F_{\1 ijk}
-\ft12 e^{\vec c\cdot\vec\phi}\, {*F_{\2}}\wedge F_{\2}\nn\\
&&-\ft12 \sum_{i} 
        e^{\vec c_{i}\cdot\vec\phi}\, {*F_{\1 i}}\wedge F_{\1 i}
-\ft12 \sum_i e^{\vec b_i\cdot\vec\phi}\, {*\cF_{\2 i}}\wedge \cF_{\2 i}
\label{d9dlag}\\
&&-\ft12 \sum_{i<j} 
        e^{\vec b_{ij}\cdot\vec\phi}\, {*\cF^i_{\1 j}}\wedge \cF^i_{\1 j}
 + {\cal L}_{\sst{FFA}}\ .\nn
\eea
%%%%%
Here, the notation for the dilatons is that $\vec\phi=(\varphi,\phi_1,
\phi_2, \ldots, \phi_{\sst{9-D}})$.

     Let us now focus on the reduction to $D=3$, and study the global
symmetry group.  There are 7 dilatons $\vec\phi$; $20+6+15=41$ axions
$A_{\0 ijk}$, $A_{\0 i}$ and $\cA^i_{\0 j}$ coming from $A_\3$, $A_\1$
and the metric; and $15+1+6=22$ vectors $A_{\1ij}$, $A_{\1 }$ and
$\cA^i_\1$ (coming from the same three sources).  We can dualise the
vectors to give further axions, whose dilaton vectors will be the
negatives of those for the vector potentials, namely $-\vec a_{ij}$,
$-\vec c$ and $-\vec b_i$.  In total, we therefore now have 7 dilatons and
63 axions.   This is precisely the dimension of the coset $E_7/SU(8)$.
As usual, the global symmetry is best understood by noting that the 
dilaton vectors for the axions of the fully-dualised three-dimensional
theory form the positive roots of $E_7$.  This is easily seen by
observing that the dilaton vectors $\vec b_{i,i+1}$ \ ($1\le i\le 5$),
$\vec a_{123}$ and $\vec c_1$ can be taken as the simple roots, and
all the other dilaton vectors can then be written as linear
combinations of these with non-negative integer coefficients.  From
the defining properties of the dilaton vectors, we see that these
simple roots give the $E_7$ Dynkin diagram:
\bigskip\bigskip
 
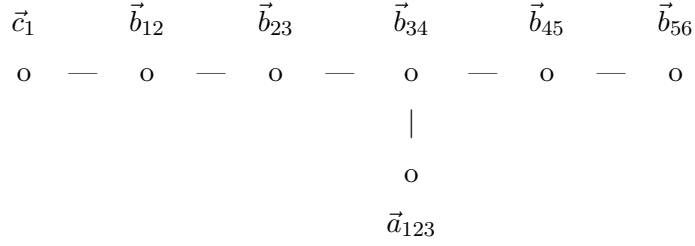
\centerline{
\begin{tabular}{ccccccccccc}\\
 $\vec c_1$& &$\vec b_{12}$& &$\vec b_{23}$& &$\vec b_{34}$& &$\vec b_{45}$
& &$\vec b_{56}$\\
 o&---&o&---&o&---&o&---&o&---&o\\
 && &   & &   &$|$&   & &   & \\
 && &   & &   &o&   & &   &  \\
 && &   & &   &$\vec a_{123}$& & &   &  \\
\end{tabular}}
\bigskip
 
\begin{itemize}
\item[]{\bf Diagram 4}: The dilaton vectors $\vec b_{i,i+1}$, $\vec
  c_1$ and $\vec a_{123}$ generate the $E_7$ Dynkin diagram. The roots
  add from left to right as one descends through the dimensions. The
  short leg starts as $\R$ in dimension $9$, and becomes $SL(2, \R)$
  in dimension $6$.
\end{itemize}
\bigskip\bigskip

   The full system of $E_7$ positive roots is then generated by
%%%%%
\bea
&&\vec b_{ij} = \vec b_{ik} + \vec b_{kj}\ ,\qquad
 \vec c_j = \vec b_{ij} + \vec c_i\ ,\nn\\
&& \vec a_{\ell jk} = \vec b_{i\ell} + \vec a_{ijk}\ ,\  \hbox{etc.}
\qquad -\vec a_{kj} =\vec b_{ik} +\vec a_{ij}\ ,\ \hbox{etc.}  \nn\\
&&-\vec a_{12} = 2\, \vec b_{12} + 3\, \vec b_{23} +3\, \vec b_{34} 
+ 2\, \vec b_{45} + \vec b_{56} + \vec a_{123} + \vec c_1\ ,\nn\\
&&-\vec c = \vec b_{12} + 2\, \vec b_{23} + 3\, \vec b_{34} + 2\, \vec
b_{45} + \vec b_{56} + 2\, \vec a_{123} \ .\label{e7sumrules}
\eea
%%%%%
It is easy to verify that these summation rules imply that all the
dilaton vectors for the axions can be expressed as linear combinations
of the simple-root dilaton vectors, with non-negative integer coefficients. 
The first four lines are trivial, implying that the fields carrying
$i,j,\ldots$ indices fall into representations under $SL(6,\R)$, which
originates from the general-coordinate transformations of the internal
6-torus. The last two lines give the non-trivial extension from
$SL(6,\R)$ to $E_7$.  

   In an intermediate dimension $3< D< \Dm=10$, we can read off the
global symmetry groups by removing the dots in the Dynkin diagram 3
whose indices lie outside the range 1 to $9- D$.  Thus the oxidation
of the $E_7$-symmetric three-dimensional scalar Lagrangian to $D$
dimensions gives theories with the following global symmetry groups:

\bigskip\bigskip
\centerline{
\begin{tabular}{|c|c|
           c|}\hline
Dim. & $G$ & $H$ \\ \hline
$\Dm=10$ & --- & ---\\ \hline
$D=9$ & $O(1,1)$ & --- \\ \hline   
$D=8$ & $GL(2,\R)$ &$ O(2)$ \\ \hline
$D=7$ & $GL(3,\R)$ & $O(3)$ \\ \hline
$D=6$ & $SL(4,\R)\times SL(2,\R)$ & $O(4)\times O(2)$ \\ \hline
$D=5$ & $SL(6,\R)$ & $O(6)$ \\ \hline
$D=4$ & $O(6,6)$ & $O(6)\times O(6)$ \\ \hline
$D=3$ & $E_7$ & $SU(8)$ \\ \hline
\end{tabular}
            }
\bigskip
\centerline{{\bf Table 1}: Scalar cosets for the $E_7$ sequence}
\bigskip\bigskip

Note that in $D\ge 7$, the global symmetry is simply the $GL(9-D,\R)$
general coordinate symmetry of the internal $(9-D)$-torus.  In $D\le
6$, the global symmetry contains a part that lies outside the general
coordinate transformations.   It should be emphasised that in all the
dimensions, the global symmetries given in Table 1 arise when all  
fields are dualised if this results in a reduction of their degrees.
Again the short leg connects at the appropriate place for a (self-dual)
4-form potential.

\subsubsection{$E_6$ sequence of scalar cosets}

The oxidation endpoint of the $E_6$ sequence is $D=8$, and again the
associated theory is the smallest one obtainable as a consistent
truncation of maximal supergravity in which the 3-form potential is
retained.  It comprises the metric, a dilaton and an axion, together
with the 3-form potential.  The eight-dimensional Lagrangian is given
by 
%%%%%
\be 
{\cal L}_8 = R\, {*\oneone} -\ft12 {*d\varphi}\wedge d\varphi
-\ft12 e^{2\varphi}\, {*d\chi\wedge d\chi} -\ft12 e^{-\varphi}\,
{*F_\4\wedge F_\4} + \chi\, dA_\3\wedge dA_\3\ ,
\label{d8trunc}
\ee
%%%%%
where $F_\4=dA_\3$.  The theory has a global $SL(2,\R)$ symmetry
at the level of its equations of motion, with $F_\4$ and its dual
$e^{-\varphi}{*F_\4}$ forming a doublet.   The details of the
Lagrangians in $D$ dimensions follow straightforwardly from the
discussion in Appendix A.  

     As in the previous case, let us now focus on the reduction to $D=3$,
and study the global symmetry group.  There are 6 dilatons $\vec\phi$;
$10+ 10 +1 =21$ axions $A_{\0 ijk}$, $\cA^i_{\0 j}$ and $\chi$; and
$10+5=15$ vectors $A_{\1ij}$ and $\cA^i_\1$.  We can dualise the
vectors to give further axions, whose dilaton vectors will be the
negatives of those for the vector potentials.  In total, we therefore
now have 6 dilatons and 36 axions, which is precisely the dimension of
the coset $E_6/U\!Sp(8)$.

     We now find that the dilaton vectors $\vec b_{i,i+1}$ \ ($1\le i\le
4$), $\vec a_{123}$ and $\vec c$ can be taken as the simple roots, and
all the other dilaton vectors can then be written as linear
combinations of these with non-negative integer coefficients.  Here,
we are using $\vec c$ to denote the dilaton vector for the axion
$\chi$.  From the defining properties of the dilaton vectors, we see
that these simple roots give the $E_6$ Dynkin diagram:
\bigskip\bigskip
 
\centerline{
\begin{tabular}{ccccccc}\\
$\vec b_{12}$& &$\vec b_{23}$& &$\vec b_{34}$& &$\vec b_{45}$\\
 o&---&o&---&o&---&o\\
  &   & &   &$|$&   &  \\
  &   & &   &o&$\vec a_{123}$&   \\
  &   & &   &$|$&   &   \\
  &   & &   &o&$\vec c$ &   \\
\end{tabular}}
\bigskip
 
\begin{itemize}
\item[]{\bf Diagram 5}: The dilaton vectors $\vec b_{i,i+1}$, $\vec c$
  and $\vec a_{123}$ generate the $E_6$ Dynkin diagram.  The roots add
  from left to right as one descends through the dimensions. In $D=5$,
  a root combines with the $\R$ factor from 7 dimensions to form a
  second $SL(3, \R)$ containing the top-dimension scalars; the group
  becomes simple below that.
\end{itemize}
\bigskip\bigskip

     We may again also read off the global symmetry groups in all
intermediate dimensions $3 < D < \Dm=8$, by making the appropriate 
truncations of the Dynkin diagram.  Thus we find for the $E_6$ sequence

\bigskip\bigskip
\centerline{
\begin{tabular}{|c|c|c|}\hline
Dim. & $G$ & $H$ \\ \hline
$\Dm=8$ & $SL(2,\R)$ &$ O(2)$ \\ \hline
$D=7$ & $GL(2,\R)$ & $O(2)$ \\ \hline
$D=6$ & $GL(2,\R)\times SL(2,\R)$ & $O(2)\times O(2)$ \\ \hline
$D=5$ & $SL(3,\R)\times SL(3,\R) $ & $O(3)\times O(3)$ \\ \hline
$D=4$ & $SL(6,\R)$ & $O(6)$ \\ \hline
$D=3$ & $E_6$ & $U\!Sp(8)$ \\ \hline
\end{tabular}}

\bigskip
\centerline{{\bf Table 2}: Scalar cosets for the $E_7$ sequence}
\bigskip\bigskip

    One notes again the Ehlers-type phenomenon of a build-up from a
trivial $\R$ symmetry to give a simple duality group in lower
dimensions.  Here, it is $SL(3,\R)$ in dimension $5$.  
\bigskip

\subsubsection{$E_n$ ($n\le5$) sequences of scalar cosets}

     The above discussion can be extended to the set of cases where we
begin in $D\le 7$ with a maximal consistent truncation of the
corresponding maximal supergravity, in which just the metric, a
dilaton and a 3-form potential are retained.  (In $D=4$, the 3-form is
non-dynamical, and will be set to zero.)  In all these cases, the
Lagrangian can be expressed as
%%%%%
\be
{\cal L}_{\sst D} = R\, {*\oneone} -\ft12 {*d\varphi}\wedge d\varphi 
-\ft12 e^{a\varphi}\, {*F_\4} \wedge F_\4\ ,\label{dtrunc}
\ee
%%%%%
where $a^2 = \ft{2(11-D)}{D-2}$.  After dimensional reduction to $D=3$
and dualisation of the vectors, the scalar coset has the global
symmetry $E_{\sst{D-2}}$, where $D\le 7$ is understood here to be the
oxidation endpoint dimension $\Dm$.  To be more precise, we have

%%%%%
\bea
\hbox{\underline{$D=3$ Scalar coset}}  && 
\hbox{\underline{Oxidation Endpoint}} \nn\\
&&\nn\\
O(5,5)/(O(5)\times O(5)) && \quad \qquad \Dm=7 \nn\\
SL(5,\R)/O(5) && \quad \qquad \Dm=6  \nn\\
SL(3,\R)\times SL(2,\R)/(O(3)\times O(2)) && \quad \qquad \Dm=5\nn\\
GL(2,\R)/ O(2) && \quad \qquad \Dm=4 \nn\\
\R && \quad \qquad \Dm=3\nn
\eea
%%%%%
for the global symmetries of the three-dimensional theories coming
from the reductions of (\ref{dtrunc}) in the indicated endpoint dimensions.

     These results can be understood very simply as follows. In the case
$\Dm=7$, the 3-form potential can be dualised to a 2-form, and hence
the Lagrangian is like the low-energy effective theory for a
seven-dimensional bosonic string.  This means that the global symmetry
group will be at least the usual $O(n,n)$ T-duality group when the
theory is compactified on an $n$-torus.  When $n=3$, the resulting
four-dimensional theory will actually have an $O(3,3)\times SL(2,\R)$
symmetry, where the extra $SL(2,\R)$ factor is the electric/magnetic
S-duality \cite{sen}.  When $n=4$, the three-dimensional theory has
$O(5,5)$ rather than simply $O(4,4)$ as its global symmetry.

    For the $E_4$ sequence, which starts in $\Dm=6$, the 3-form
potential is dual to a vector, and in fact the dualised theory can be
obtained by dimensional reduction from pure gravity in $D=7$.  Thus
the global symmetry of the three-dimensional theory will be the same
as that for pure gravity reduced from seven dimensions, namely
$SL(5,\R)\sim E_4$.  Again, this is in accordance with the
$SL(D-2,\R)$ rule mentioned in the Introduction.

   For the $E_3$ sequence, starting in $\Dm=5$, the 3-form is dual to
an axion, and hence the five-dimensional theory comprises the metric
and a dilaton-axion system with an $SL(2,\R)$ symmetry.  Upon
reduction to three dimensions, the gravity sector contributes an
$SL(3,\R)$ factor, giving $SL(3,\R)\times SL(2,\R)\sim E_3$ in total.

  For the $E_2$ sequence, beginning in $\Dm=4$, we have just
pure gravity plus a dilaton.  The global symmetry of the reduced
three-dimensional theory is therefore $GL(2,\R)\sim E_2$, where the
$SL(2,\R)$ factor is the usual one from the reduction of gravity to
$D=3$, and the extra $\R$ factor is from the shift symmetry of the
four-dimensional dilaton.  

   Finally, for the $E_1\sim \R$ sequence, we simply have gravity plus
a scalar in $D=3$, and so no oxidation is possible.

\subsubsection{Summary}

    It is well known that the dimensional reduction of eleven-dimensional
supergravity to $D=3$ gives rise to a scalar coset theory with an
$E_8$ global symmetry.  In other words, the oxidation endpoint of the
three-dimensional $E_8/SO(16)$ scalar coset is $D=11$ supergravity.
In this section we studied the oxidation endpoints of the
three-dimensional scalar cosets whose global symmetry groups are the
$E_n$ subgroups of $E_8$, with $2\le n\le 8$.  We showed that in the
oxidation endpoint dimension for the subgroup $E_n$ is generically
given by $\Dm=n+2$.  In all cases, the endpoint theory includes the
metric, a dilaton, and a 3-form potential, and in $\Dm\le 7$ there are
no additional fields.  In $ D = 10$, corresponding to $E_8$, there is
also a 2-form potential and a vector potential.  In $D=9$,
corresponding to $E_7$, there is instead just an additional vector
potential.  In $D=8$, there is instead an additional 0-form potential,
or axion.

   There are, as we saw, three special cases that arise.  For $E_8$ the
``endpoint'' implied by the generic discussion, namely $D=10$, can be
further oxidised to the bosonic sector of $D=11$ supergravity.  For
$E_7$, the generic discussion leads to an endpoint in $D=9$, but
again a further oxidation is possible, giving, in this case, the
truncation of type IIB supergravity in $D=10$ to the metric plus the
self-dual 5-form.  The third special case is $E_4$, for which the
``endpoint'' in $D=6$ can be further oxidised to pure gravity in
$D=7$, after first dualising the 3-form potential to a vector in
$D=6$.  These are the only three cases among the $E_n$ groups that
contain $SL(n+1,\R)$ as subgroups, which explains why in each case a
further oxidation by one step is possible.  (A $D$-dimensional theory
involving gravity, when compactified to three dimensions, is expected
to exhibit an $SL(D-2,\R)$ global symmetry.)

    Our results for the oxidation sequences for the various $E_n$
cosets in $D=3$ may be presented in the following table, which serves to make
apparent a rather intriguing symmetry.  The oxidation sequence for
each of the three-dimensional $E_n$ cosets is presented vertically,
with $n$ plotted horizontally (as usual, $E_5\sim D_5$, $E_4\sim A_4$,
$E_3\sim A_1\times A_2$, $E_2\sim \R\times A_1$, $E_1\sim \R$, and
$E_0$ is trivial):

\bigskip\bigskip
\centerline{
\begin{tabular}{|c|c|c|c|c|c|c|c|c|c|}\hline
$D=11$ & - & & & & & & & &\\ \hline
$D=10$ &$\R$ &  &&&&&&& \\ \hline  
$D=9$ & $\R\times A_1$ & $\R$ & & &&&&& \\ \hline
$D=8$ & $A_1\times A_2$ & $\R \times A_1$ & $A_1$ & &&&&&\\ \hline
$D=7$ & $E_4$ & $\R \times A_2$ & $\R \times A_1$ & $\R$ & - & & &&\\ \hline
$D=6$ & $E_5$ & $ A_1\times A_3$ & $\R\times A_1^2$
        & $ \R ^2$ & $\R$ & & & &    \\ \hline
$D=5$ & $E_6$ & $A_5$ & $A_2^2$ & $
       \R\times A_1^2$ & $\R \times A_1$ & $A_1$ & & &\\ \hline
$D=4$ & $E_7$ & $D_6$ & $A_5$ & $A_1\times A_3$ &
        $\R \times A_2$ & $\R \times A_1$ & $\R$ & &  \\ \hline
$D=3$ & $E_8$ & $E_7$ & $E_6$ & $E_5$ & $E_4$ & $A_1\times A_2$ & $\R\times 
A_1$ & $\R$&-\\ \hline
\end{tabular}
            }
\bigskip
\centerline{{\bf Table 3}: Disintegration (\ie Oxidation) for $E_n$ Cosets}
\bigskip\bigskip

%   We have written the groups in each entry here in the fashion in
%which they naturally arose in the previous discussion.  Bearing in
%mind, however, that we have the isomorphisms $O(1,1)\sim \R$,
%$O(2,2)\sim SL(2,\R)\times SL(2,\R)$, $O(3,3)\sim SL(4,\R)$ and
%$GL(2,\R)\sim SL(2,\R)\times \R$, together with the usual $E_1\sim
%\R$, $E_2\sim GL(2,\R)$, $E_3\sim SL(3,\R)\times SL(2,\R)$, $E_4\sim
%SL(5,\R)$ and $E_5\sim O(5,5)$, we see that there is a 
%reflection symmetry across the diagonal line linking $E_8$, $O(6,6)$,
%$SL(3,\R)\times SL(3,\R)$ and $O(1,1)\times \R$. The column above $E_5=O(5,5)$
%is made of orthogonal groups in fact.

     Note that in each sequence of theories represented by a vertical
column in Table 3, the sequence of symmetry groups is obtained by
successively deleting the vertices of the $D=3$ Dynkin diagram as one
ascends through the dimensions.  In particular, in the step from $D=3$
to $D=4$, the vertex of $E_n$ (with $n\ge4$) that is deleted is always
the one that connects to the ``extra'' vertex of the extended Dynkin
diagram for $E_n$.   Note also that an entry ``$-$'' 
indicates that the sequence has an oxidation endpoint with no
global symmetry in that dimension.  (We have not included the
ten-dimensional ancestor of $E_7$, nor the seven-dimensional ancestor
of $E_4$.)

  Most striking is the ``magic'' reflection symmetry across the
diagonal, which is much more complete than the approximate symmetry
mentioned in \cite{b85} for the non-maximally non-compact duality
groups of $N=4$ pure supergravities.  This might lead to a more
systematic and deeper understanding of U-dualities, since we now know
of three ``magic tables.'' As a comment we should point out that the
fourth slot of the magic square of Tits-Rosenfeld-Freudenthal is
absent; this does not seem to follow from a special choice of real
forms but rather it suggests that a new geometry remains to be
discovered here.

\section{Non-simply-laced scalar coset sequences\label{sec:nonsimple}}

    In all the examples in the previous section, the global symmetry
groups of the three-dimensional scalar cosets were simply-laced.  There
are further examples in the classification of \cite{bgm} for which the
three-dimensional coset manifold has a non-simply-laced symmetry
group, namely $B_n$, $C_n$, $G_2$ and $F_4$.  Here, we shall consider
the cases where the group is maximally non-compact.

\subsection{$B_n$ sequences}

    The maximally non-compact form of $B_n$ is $O(n+1,n)$.  The $D=3$
scalar coset manifold for this case can be obtained by starting in
$\Dm=n+2$ dimensions with a theory comprising the  metric, a dilaton, a
2-form potential and a vector potential.  The Lagrangian is given by
%%%%%
\be
{\cal L} = R\, {*\oneone}  -\ft12{*d\varphi}\wedge d\varphi - \ft12
e^{a\varphi}\, {*F_\3}\wedge F_\3 - \ft12 e^{\fft12a\varphi}\, {*F_\2}
\wedge F_\2\ , \label{bnlag}
\ee
%%%%%
where $a^2= 8/(\Dm-2)$, and
%%%%%
\be
F_\3 = dA_\2 + \ft12 A_\1\wedge dA_\1\ ,\qquad F_\2 = dA_\1\ .
\ee
%%%%%
(This is really a special
case of the standard type of construction in string theory:
If one starts instead with $N$ vectors in $\wtd D$ dimensions, one gets an  
$O(\wtd D-D + N,\wtd D -D)$ symmetry in $D$ dimensions, enlarging to
$O(\wtd D -2 + N, \wtd D-2)$ in $D=3$ \cite{sen2}.)

   Note that the special (stringy?) case of 1 vector added to a bosonic type I
sector in dimension 10 would possibly lead to an hyperbolic algebra of 
maximal rank 10, very much like the conjectured $E_{10}$ symmetry
or the overextended $D_8$ \cite{b85} of the type II and I theories. For a 
classification of hyperbolic Coxeter groups see, for instance, \cite{H}.

    An interesting special case arises when $n=3$.  The oxidation
endpoint following the above discussion would be in $\Dm = 5$.  In
fact, this theory can be oxidised further, to $\Dm = 6$, to $N=1$
supergravity, whose bosonic sector comprises the metric and a
self-dual 3-form.  The dimensional reduction of this theory to $D=5$
and $D=4$ was discussed in detail in \cite{ugen}.  In particular, in
$D=4$, the global symmetry group is $O(2,2)$.  When it is further
reduced to $D=3$, the global symmetry becomes $O(4,3)$.  We use the
standard notation, where we take $\vec a$ to denote the dilaton
vectors for the fields coming from the reduction of the self-dual
3-form, and $\vec b$ for those coming from the metric.  In $D=3$,
there are a total of 3 dilatons and 9 axions, after dualising the
three Kaluza-Klein vectors.  We can verify that the 9 axion dilaton
vectors form the full set of positive roots of $O(4,3)$, with $\vec
b_{12}$, $\vec b_{23}$ and $\vec a_{12}$ being the simple roots:
\bigskip\bigskip
 
\centerline{
\begin{tabular}{ccccccc}\\
$\vec b_{12}$& &$\vec b_{23}$& &$\vec a_{12}$\\
 o&---&o&$=\!=\!=$&$\bullet$\\
\end{tabular}
}
\bigskip

\begin{itemize}
\item[]{\bf Diagram 6}: The dilaton vectors $\vec b_{12}$, $\vec
b_{23}$ and $\vec a_{12}$ generate the $B_3$ Dynkin diagram, with the
black dot indicating the shorter root.
The 2-form potential leads to a branching off here of the shorter root.
\end{itemize}

\bigskip\bigskip

   The reason why the further oxidation to $\Dm=6$ is possible in this
case is that the $O(4,3)$ group contains $SL(4,\R)$ as a subgroup.
This $SL(4,\R)$ is associated with the global symmetry already present
in any theory involving gravity, indicating an oxidation endpoint in $\Dm=6$.
 
 For higher $n$'s the growth  of the linear subgroup of $B_n$ is away from the 
small root.
 
\subsection{$C_n$ sequences}

     The maximally non-compact form of $C_n$ is $Sp(2n,\R)$, and the
corresponding three- dimensional scalar cosets are $Sp(2n,\R)/U(n)$.
These cosets are different from the others that we have discussed, in
that their oxidation endpoints are all in four dimensions, and no
further oxidations beyond $D=4$ are possible.  The endpoints in $\Dm =
4$ have the global symmetry $Sp(2n-2,\R)$ \cite{bgm}.

   To see this, let $\vec \sigma_\a$ be the $(n-1)$-component
positive-root vectors of $Sp(2n-2,\R)$.  These can be written in terms
of a basis of $(n-1)$ unit vectors $\vec e_i$ as
%%%%%
\be
\vec e_i \pm \vec e_j\ ,\quad i>j\ ,\qquad {\rm and}\qquad 2\, \vec e_i
\ ,
\ee
%%%%%
where we are defining positivity in terms of the sign of the first
non-vanishing component starting from the right.  The simple roots are
%%%%%
\be
\vec e_{i+1} - \vec e_i\ ,\qquad {\rm and}\qquad 2\, \vec e_{1}\ .
\ee
%%%%%
The full set of positive roots for $Sp(2n,\R)$ is then given by the
$n$-component vectors
%%%%%
\be
(\vec \sigma_\a,0)\ ,\qquad (\pm\vec e_i, 1)\ ,\qquad (\vec 0,2)\ .
\label{cnpos}
\ee
%%%%%
These have a simple Kaluza-Klein interpretation as a reduction from
$D=4$ to $D=3$.  The root $ (\vec 0,2)$ arises as the dilaton vector
associated with the dual of the Kaluza-Klein vector, while the roots
$(\vec \sigma_\a,0)$ originate from the dilaton vectors of the axions
in an $Sp(2n-2,\R)/U(n-1)$ scalar coset in $D=4$, with $\vec
\sigma_\a$ as positive root vectors.  The roots $(\vec e_i, 1)$ arise
as the dilaton vectors for the direct dimensional reductions of a set
of $(n-1)$ vector potentials $A_\1^i$, to give axions $A_\0^i$ in
$D=3$.  In addition, the vector potentials will also give rise to
vector potentials in $D=3$, and after dualising these to axions, their
associated dilaton vectors will be $(-\vec e_i,1)$, accounting for the
remaining root vectors in (\ref{cnpos}).  Thus the $Sp(2n,\R)/ U(n)$
sigma model in $D=3$ oxidises to give the Lagrangian
%%%%%
\be
{\cal L}_4 = -\ft12{*\del\vec\phi}\wedge d\vec\phi 
 -\ft12\sum_\a e^{\vec \sigma_\a \cdot
\vec\phi}\, {*G_\1^\a}\wedge G_\1^a
 - \ft12 \sum_{i=1}^{n-1} e^{\vec e_i\cdot\vec 
\phi_i}\, {*F_\2^i}\wedge F_\2^i\label{cnd4lag}
\ee
%%%%%
in $D=4$.  Here, the $(n-1)$ dilatons $\vec\phi=(\phi_1,\cdots,\phi_{n-1})$
are associated with the Cartan subalgebra of $Sp(2n-2,\R)$, and the $\ft12
n(n-1)$ 1-form field strengths $G_\1^\a$ are formed from the
axions associated with the positive roots of $Sp(2n-2,\R)$. The
vector potentials $A_\1^i$ transform under the fundamental
representation of $Sp(2n-2,\R)$.

     It is easily seen that no further oxidation beyond $\Dm=4$ is
possible for this $C_n$ sequence of three-dimensional scalar cosets.
One way to see this is to note that the field strengths $F_\2^i$ in
(\ref{cnd4lag}) have dilaton vectors $\vec e_i$ that all have
(length)$^2=1$.  If (\ref{cnd4lag}) were to come from the Kaluza-Klein
reduction of a theory in $D=5$, then the Kaluza-Klein vector from this
reduction step would have to be interpreted as one of the vector
potentials in $D=4$.  This is not possible, however, since the
Kaluza-Klein vector would have a dilaton vector with (length)$^2=3$,
rather than (length)$^2=1$.  Thus $\Dm=4$ is the oxidation endpoint
for the $C_n$ sequence of three-dimensional scalar cosets.

We may recall here that the affine extension of the group symmetry
after reduction to 2 dimensions includes a manifest $SL(\Dm-2, \R)$
symmetry. A look at the Dynkin diagram of the affine $C^{1}_n$
immediately shows that $\Dm=4$, since the linear subgroup should start
from one of the long root vertices.

\subsection{$G_2$ sequence}

   There is a $G_2/(SU(2)\times SU(2))$ maximally non-compact scalar
coset theory in $D=3$ \cite{bgm}.  We find that it can be oxidised
back to an endpoint in $\Dm=5$, where it becomes the bosonic sector
of simple supergravity.  This is the Einstein-Maxwell
system in $D=5$, with an $FFA$ term:  
%%%%%
\be
{\cal L}_5 = R\, {*\oneone} - \ft12 {*F_\2}\wedge F_\2
+\ft1{3\sqrt3} F_\2\wedge F_\2\wedge A_\1\ .\label{g2lag}
\ee
%%%%% 
Upon reduction to $D=3$, we obtain, in the standard notation, the
Lagrangian
%%%%%
\bea
{\cal L} &=& R\, {*\oneone} -\ft12 {*d\vec\phi}\wedge d\vec\phi -
\ft12 e^{\phi_2 -\sqrt3\phi_1}\, {*\cF^1_{\1 2}}\wedge \cF^1_{\1 2}
-\ft12 e^{\fft2{\sqrt3}\phi_1}\, {*F_{\1 1}}\wedge F_{\1 1} \nn\\
&&-\ft12 e^{\phi_2 -\ft1{\sqrt3}\phi_1}\,  {*F_{\1 2}}\wedge F_{\1 2} 
- \ft12 e^{-\phi_2 -\sqrt3\phi_1}\, {*\cF_\2^1}\wedge \cF_\2^1
\label{d5einstmax}\\
&&- \ft12 e^{-2\phi_2}\, {*\cF_\2^2}\wedge \cF_\2^2
-\ft12 e^{-\phi_2 -\ft1{\sqrt3}\phi_1}\, {*F_\2}\wedge F_\2
+ \ft{2}{\sqrt3} dA_{\0 i}\wedge dA_{\0 2}\wedge A_\1\ .\nn
\eea
%%%%%
After dualising the vector potentials to give axions, we see that
there will be six axions, together with the two dilatons.  We find
that we may take the dilaton vectors $\vec\a_1=(-\sqrt3,1)$ and
$\vec\a_2=(\ft2{\sqrt3}, 0)$, corresponding to the axions $\cA^1_{\0
2}$ and $A_{\0 1}$, as the simple roots of $G_2$, with the remaining
dilaton vectors expressed in terms of these as
%%%%%
\be
(-\ft1{\sqrt3}, 1)=\vec\a_1+\vec \a_2\ ,\quad (\ft1{\sqrt3},1)=
\vec\a_1 + 2\vec\a_2\ ,\quad
(\sqrt3,1)= \vec\a_1+3\vec\a_2\ ,\quad
(0,2)=2\vec\a_1+3\vec a_3\ .
\ee
%%%%%
It is easily verified that $\vec\a_1$ and $\vec\a_2$ generate the
Dynkin diagram for $G_2$:
\be
{\rm o}\equiv\!\equiv\!\equiv \bullet\ .
\ee

    It is interesting to look also at the global symmetry of the
oxidation of the $G_2$ coset to $D=4$.  From the $G_2$ Dynkin diagram,
we expect that there should be an $SL(2,\R)$ symmetry.  The $D=4$
Lagrangian is 
%%%%%
\bea
{\cal L}_4 &=& R\, {*\oneone} -\ft12{*d\phi_1}\wedge d\phi_1 - \ft12
e^{\ft2{\sqrt3}\phi_1}\, {*F_{\1 1}}\wedge F_{\1 1} -\ft12 e^{-\sqrt3
  \phi_1}\, {*\cF^1_{\2}}\wedge \cF^1_\2 \nn\\
&&-\ft12 e^{-\ft1{\sqrt3}\phi_1}\, 
{*F_\2}\wedge F_\2 +\ft1{\sqrt3} A_\0\, dA_\1\wedge dA_\1
\ .
\eea
It can be seen from this that the scalar manifold, comprising the
dilaton $\phi_1$ and the axion $A_{\0 1}$, indeed has an $SL(2,\R)$
symmetry.  It can also be seen that the 1-component dilaton vectors
for the two vector potentials, together with their negatives,
form the weights of the 4-dimensional representation of $SL(2,\R)$.

     A discussion of the dimensional reduction of simple supergravity 
in $D=5$, and its global symmetry, was also given in \cite{mo}.

\subsection{$F_4$ sequence}

   We find that the oxidation endpoint of the $F_4$ sequence is a $\Dm=6$
dimensional theory, containing the metric, a dilaton, an axion, two
vectors, a 2-form potential and a self-dual 3-form field strength.
The Lagrangian is given by
%%%%%%%%%%%%
\bea
{\cal L}_6 &=& R\, {*\oneone} -\ft12{*d\varphi}\wedge d\varphi -\ft12
e^{\sqrt2\varphi}\, {*d\chi}\wedge d\chi -\ft12 e^{-\sqrt2\varphi}\,
{*F_\3}\wedge F_\3\nn\\
&& -\ft12 {*G_\3}\wedge G_\3 -\ft12 e^{\ft1{\sqrt2}\varphi}\, {*F_\2^+}
\wedge F_\2^+ -\ft12e^{-\ft1{\sqrt2}\varphi}\, {*F_\2^-}\wedge {F_\2^-}
\label{d6sl2r}\\
&&-\ft1{\sqrt2} \chi\, F_\3\wedge G_\3 - \ft12 A_\1^+\wedge
F_\2^+\wedge F_\3 -\ft12 A_\1^+\wedge F_\2^-\wedge G_\3\ .\nn
\eea
%%%%%%%%%
The self-duality condition on the 3-form $G_\3$ is to be imposed after
having obtained the equations of motion.  
The field strengths are given in terms of potentials as follows:
%%%%%
\bea
&&F_\3 = dA_\2 +\ft12 A_\1^-\wedge dA_\1^-\ ,\qquad
G_\3 = dB_\2 -\ft1{\sqrt2} \chi\, F_\3 -\ft12 A_\1^+\wedge dA_\1^-\ ,\nn\\
&&F_\2^+ = dA_\1^+ + \ft1{\sqrt2}\chi\, dA_\1^-\ ,\qquad F_\2^- = dA_\1^-\ .
\label{d6fields}
\eea
%%%%%
It is easily verified that
the self-duality constraint is consistent with the equations of motion
and Bianchi identities.  It can also be seen
that the theory described by (\ref{d6sl2r}) has a global $SL(2,\R)$
symmetry at the level of the equations of motion, with the two vectors
$A_\1^+$ and $A_\1^-$ forming a doublet, and the 3-form fields $F_\3$,
${*F_\3}$ and $G_\3$ forming the $(+,-,2)$ components of a triplet.
As we shall explain in section 3.5, the Lagrangian (\ref{d6sl2r}) can
be obtained by starting with the eight-dimensional 
Lagrangian (\ref{d8trunc}) which is the oxidation endpoint of the
$E_6$ sequence, and reducing to $D=6$.  After performing the
reduction, the six-dimensional Lagrangian has a $GL(2,\R)\times
SL(2,\R)$ global symmetry.  A consistent truncation can then be
performed, under which a diagonal $SL(2,\R)$ survives; this truncated
theory is described by the Lagrangian (\ref{d6sl2r}) together with the
self-duality constraint $G_\3={*G_\3}$.   This six-dimensional theory
is the bosonic sector of a supersymmetric $N=1$ chiral supergravity,
with two anti-self-dual tensor multiplets and two vector multiplets.  This
is a similar to the class of chiral six-dimensional
supergravities constructed in \cite{sagnotti}; see Appendix B for a
detailed discussion.

     In the standard notation, and associating dilaton vectors $\vec
a$, $\vec e$, $\vec c^{\,+}$, $\vec c^{\,-}$, $\vec b$ and $\vec h$
with the fields $F_\3$, $G_\3$, $F_\2^+$, $F_\2^-$, $\cF$ and $\chi$
respectively, we find, after reduction to $D=3$ and dualisation of the
vectors, that the dilaton vectors of all the 24 axions form the
positive roots of the exceptional Lie algebra $F_4$.  In particular,
$\vec b_{23}$, $\vec b_{12}$, $\vec c^{\,+}_1$ and $\vec h$ are the
positive simple roots, giving
\bigskip\bigskip
 
\centerline{
\begin{tabular}{ccccccccc}\\
$\vec b_{23}$& &$\vec b_{12}$& &$\vec c^{\,+}_1$& &$\vec h$ \\
 o&---&o&$=\!=\!=$&$\bullet$&---&$\bullet$\\
\end{tabular}
}
\bigskip

\begin{itemize}
\item[]{\bf Diagram 7}: The dilaton vectors $\vec b_{23}$, $\vec b_{12}$,
$\vec c^{\,+}_1$ and $\vec h$ generate the $F_4$ Dynkin diagram. 
The build-up of the symmetry is now towards the left end.
\end{itemize}

\bigskip\bigskip

 From this Dynkin diagram, it is straightforward to see that the
oxidations of the three-dimensional $F_4$ theory to $D=4$ and $D=5$
will have global symmetries $Sp(6,\R)$ and $SL(3,\R)$ respectively.

\subsection{Embedding in simply-laced cosets}

    The non-simply-laced algebras can all be interpreted as embeddings
in larger simply-laced algebras.  At the level of their Dynkin
diagrams, these embeddings can be viewed as ``folding over'' the
diagrams for the simply-laced diagrams, so that two or more vertices
become identified.  These are described in terms of Satake diagrams in
\cite{helg}.  These identifications in turn have an interpretation at
the level of the coset Lagrangians, as consistent truncations of the
Lagrangians for the simply-laced groups to give the Lagrangians for
the non-simply-laced groups.  We shall discuss the various cases
below.

\subsubsection{$B_{n-1}$ embedded in $D_{n}$}

    One can obtain the $B_{n-1}$ Dynkin diagram from the $D_{n}$
Dynkin diagram, given in Diagram 2, by folding over and identifying
the vertices $\vec b_{12}$ and $\vec a_{12}$.  This implies the
existence of a consistent truncation of the three-dimensional
Lagrangian (\ref{dnlag2}) in which the axions $\cA^1_{\0 i}$ are
equated to $A_{\0 1i}$, and the axion $\chi_1$ is equated to
$\psi^1$.  In fact the easiest way to see this is to look at the
Lagrangian in $D=n+1$, obtained by performing a single circle
reduction of the oxidation endpoint Lagrangian (\ref{dnlag}) for the
$D_n$ coset.  In $D=n+1$, we obtain the Lagrangian
%%%%%
\bea
{\cal L} &=& R\, {*\oneone} -\ft12 {*d\vec\phi}\wedge d\vec\phi -
\ft12 e^{\vec a_1\cdot\vec\phi}\, {*F_{\2 1}}\wedge F_{\2 1}\nn\\
&& -\ft12 e^{\vec b_1\cdot\vec\phi}\, {*\cF_{\2}^1}\wedge \cF_{\2}^1 
-\ft12 e^{\vec a\cdot\vec\phi}\, {*F_\3}\wedge F_\3\ .\label{dnm1lag}
\eea
%%%%%
It is easily seen that the two 2-form field strengths $F_{\2 1}$ and
$\cF_\2^1$ can be equated provided that the two dilatons in $\vec\phi$
are also truncated to a single dilaton combination, such that $(\vec
a_1-\vec b_1)\cdot\vec\phi=0$.  Defining 
%%%%%
\be
a=\ft12|\vec a_1+\vec b_1|\ ,\qquad  
\qquad \ft12(\vec a_1+\vec b_1)\cdot\vec\phi = a\, \varphi\ ,
\qquad F_\2 = \sqrt2 F_{\2 1} =\sqrt2\, \cF_\2^1\ ,
\ee
%%%%%
it follows that (\ref{dnm1lag}) can be consistently truncated to give
(\ref{bnlag}), in $D=n+1$.  This is indeed the Lagrangian that
describes the oxidation endpoint of the $B_{n-1}$ scalar coset in $D=3$.
It is easily seen that the identification of the two 1-form potentials
in $D=n+1$ implies precisely the set of identifications of axions
given above equation (\ref{dnm1lag}).

\subsubsection{$G_2$ embedded in $D_4$}

    The Dynkin diagram for $G_2$ can be obtained from that of $D_4$,
by identifying the three vertices corresponding to the three mutually 
orthogonal simple roots.  As in the $B_{n-1}$ example above, the
associated consistent truncation can best be described in the endpoint
dimension of the non-simply-laced coset, namely $\Dm=5$ for this $G_2$
case.  The endpoint dimension for the $D_4$ coset is $\Dm=6$, where
there is the metric, a dilaton and a 2-form potential, as described by
(\ref{dnlag}).  Upon reduction to $D=5$, this theory can be
consistently truncated to the bosonic sector of $D=5$ simple
supergravity, with Lagrangian given by (\ref{g2lag}).  This is
achieved by equating the three 2-form field strengths $\cF_\2^1$,
$F_{\2 1}$ and ${*F_\3}$.  It is easy to see that from the
three-dimensional viewpoint, these identifications lead to
identifications of the axions that imply precisely the identification
of the three simple roots $\vec b_{12}$, $\vec b_{34}$ and $\vec
a_{12}$ that form the ``ears'' of the $D_4$ Dynkin diagram in Diagram
2.

\subsubsection{$F_4$ embedded in $E_6$}

    This embedding corresponds to an identification of the pair of
simple roots $\vec b_{12}$ and $\vec c$, and the pair of simple roots
$\vec b_{23}$ and $\vec a_{123}$, of the $E_6$ algebra given in
Diagram 5.  The corresponding consistent truncation of the Lagrangian
can be most conveniently performed in the oxidation endpoint dimension
$\Dm=6$ of the three-dimensional $F_4$ scalar Lagrangian.  On the
other hand the endpoint of the $E_6$ sequence, as discussed in section
2.3.2, is in $\Dm=8$.  Dimensionally reducing the Lagrangian
(\ref{d8trunc}) to $D=6$, we can make a consistent truncation to give
the Lagrangian (\ref{d6sl2r}), by equating the axions $\chi$ and
$\cA^1_{\0 2}$, and the 2-form fields $\cF_\2^2$ and $F_{\2 12}$. In
this truncation, the dual of the 
4-form field strength $F_\4$ is set equal to $\cF_\2^1$, and the field
$F_{\3 2}$ is required to be self-dual.  It
is easy to see that the consequent identifications of fields in $D=3$
imply precisely the identification of simple roots listed above,
describing the embedding of $F_4$ in $E_6$.

\subsubsection{$C_n$ embedded in $A_{2n-1}$}

    This embedding is described by folding over the Dynkin diagram of
$A_{2n-1}$ at its midpoint, so that $(n-1)$ pairs of simple roots are
identified.  As usual, we make the corresponding consistent truncation
in the Lagrangian in the oxidation endpoint dimension, namely $\Dm=4$
for this $C_n$ sequence.  The oxidation endpoint for the $A_{2n-1}$
sequence is pure gravity in $\Dm=2n+2$.  Dimensionally reducing this
to $D=4$, we can make a consistent truncation given by
%%%%%
\be
 \cF_\2^ i = {* \cF_\2^{2n-1-i}}\ ,\qquad \cF^i_{\1 j} =
 \cF^{2n-1-j}_{\1 2n-1-i}\ .
\ee
%%%%%
This truncation gives precisely the Lagrangian (\ref{cnd4lag}).  The
dilaton vectors $\vec e_i$ in (\ref{cnd4lag}) are given by $\vec e_i =
\vec b_i - \vec b_{2n-1-i}$, and they are indeed orthonormal.  Reducing
one step further to $D=3$, we can see that the consequent
identifications of axions imply the necessary identification rules for
obtaining the Dynkin diagram of $C_n$ from that of $A_{2n-1}$.

\section{Non-maximally non-compact cosets}

    The discussion of three-dimensional $G/H$ scalar coset theories
where $G$ is not maximally non-compact becomes somewhat involved, and
we shall not attempt a complete classification of their oxidation
endpoints here.  The principal examples, from our point of view, are
the sets of cosets $O(p,q)/(O(p)\times O(q))$, $SU(p,q)/S(U(p)\times
U(q))$ and $SO^*(2n)/U(n)$.  The first of these sets is easily
understood from a higher-dimensional standpoint.  It is well known
that a ten-dimensional $N=1$ string with $m$ vector potentials gives
rise upon toroidal dimensional reduction to a theory with
$O(10+m-D,10-D)$ global symmetry in $5\le D\le 10$ dimensions,
$O(6+m,6)\times SL(2,\R)$ in $D=4$, and $O(8+m,8)$ in $D=3$.  This
generalises straightforwardly to other starting dimensions.  Thus if
we suppose, without loss of generality, that $p>q$, we may oxidise the
three-dimensional $O(q+m,q)/(O(q+m)\times O(q))$ scalar coset to $\Dm=
q+2$ dimensions, where it corresponds to a theory similar to that of
an $N=1$ string, comprising the metric, a dilaton, a 2-form potential
and $m$ 1-form potentials $A_\1^i$, with the Lagrangian
%%%%%
\be
{\cal L} = R\, {*\oneone} -\ft12{*d\varphi}\wedge d\varphi - \ft12
e^{a\varphi}\, {*F_\3}\wedge F_\3 - \ft12 e^{\fft12a\varphi}\, \sum_i 
{*F^i_\2}\wedge F_\2^i\ ,
\label{bnlag2}
\ee
%%%%%
where $a^2= 8/(\Dm-2)$, and
%%%%%
\be
F_\3 = dA_\2 + \ft12 A^i_\1\wedge dA^i_\1\ ,\qquad F^i_\2 = dA^i_\1\ .
\ee
%%%%%

    Other non-maximally non-compact cosets, including some further
isolated examples, are discussed in \cite{bgm}, where their oxidations
to $D=4$ are given.  It may be that these cases, like the $C_n$
sequences discussed in section 3.2, cannot be further oxidised beyond
four dimensions, but we have not investigated this in detail.  One
result is that all cosets that are Hermitean symmetric spaces
have oxidation endpoints in $D=4$. Another observation is that all
quaternionic spaces technically homogeneous under the Borel subgroup
(also called Alekseevskian spaces) have maximal oxidation
dimension 6.  For a discussion of the relation of these spaces to 
supergravities, see for instance, \cite{PW}.

  One example among those conjectured in \cite{b81,b85} was
constructed recently as an anomalous but classically supersymmetric
theory, in \cite{FK}.

\section{Summary and conclusions}

   We have examined $G/H$ scalar coset Lagrangians in three
dimensions, for all the cases where $G$ is a simple
maximally-non-compact group, and $H$ is its maximal compact subgroup.
In all cases, the three-dimensional scalar Lagrangian can be
re-interpreted as coming from the toroidal dimensional reduction of
some higher-dimensional theory comprising gravity, possibly with
additional antisymmetric tensor fields and dilatons.  Thus we have
introduced the notion of the ``oxidation endpoint'' of a
three-dimensional scalar coset Lagrangian, by which we mean the theory
in the highest possible dimension whose toroidal dimensional reduction
yields (after dualisation of all vector potentials to axions)
precisely the originally-given three-dimensional theory.  In many
cases the fields of the three-dimensional theory can be produced
rather ``economically'' from the higher dimension, in that many
three-dimensional fields are generated from each of a few
higher-dimensional ones.  By contrast there are other cases, most
notably the $Sp(2n,\R)/U(n)$ three-dimensional scalar cosets, where
less advantage is gained by oxidation.  For example, the
$Sp(2n,\R)/U(n)$ theories can be oxidised only as far as $D=4$.

    We may summarise the oxidation endpoint dimensions for all the
simple maximally-non-compact three-dimensional $G/H$ cosets in the following
table.  Complete details, together with any particular discussion for
special cases, may be found in the earlier sections.
(Note that for the $Sp(2n,\R)/U(n)$ cosets the field contents of the
$\Dm=4$ theory includes $(n-1)$ dilatons $\vec\phi$, $\ft12n(n-1)$
axions $\chi^\a$, and $(n-1)$ 1-forms $A_\1^i$, as described in
section 3.2.)

\bigskip\bigskip
\centerline{
\begin{tabular}{|c|c|c|c|}\hline
$G$  & $H$ & $\Dm$ & Fields \\ \hline
$SL(n+1,\R)$ & $O(n+1)$ &$ n+3 $& $g_{\mu\nu}$ \\ \hline
$O(n,n)$ & $O(n)\times O(n)$ & $n+2$ & $g_{\mu\nu},\ \varphi,\ A_\2$ \\
\hline
$E_8$ & $O(16)$ & 11 & $g_{\mu\nu},\ A_\3$ \\ \hline
$E_7$ & $SU(8)$ & 10 & $g_{\mu\nu},\ B_\4(\hbox{self-dual})$ \\ \hline
$E_6$ & $USp(8)$ & 8 & $g_{\mu\nu},\ \varphi,\ \chi,\ A_\3$ \\ \hline
\hline
$O(n+1,n)$ & $O(n+1)\times O(n)$ & $n+2$ & $g_{\mu\nu},\ \varphi,\
A_\1,\ A_\2$ \\ \hline
$Sp(2n,\R)$ & $U(n)$ & 4 & $g_{\mu\nu},\ \vec\phi,\ \chi^\a,\
A_\1^i$\\ \hline
$G_2$ & $SU(2)\times SU(2)$ & 5 & $g_{\mu\nu},\ A_\1$ \\ \hline
$F_4$ & $USp(6)\times SU(2)$ & 6 & $g_{\mu\nu},\ \phi,\ 
\chi,\ A_\1^+,\ A_\1^-,\ A_\2,\ B_\2(
\hbox{self-dual})$ \\ \hline
\end{tabular} }
\bigskip

\centerline{{\bf Table 4}: Oxidation endpoints for maximally-non-compact
cosets}
\bigskip\bigskip
 
  A compact way to summarise the build-up of hidden dimensions is that
$E_8, E_7, G_2$, $B_3$ and $A_n$ can be lifted to a dimension equal to
their rank plus three, $E_6, F_4, D_n$ and $B_n$ can be lifted to
their rank plus 2 dimensions, and finally $C_n$ can be lifted to
dimension 4.  The oxidation endpoints of the $A_1$, $G_2$, $B_3$,
$F_4$, $E_8$, $D_8$ and $B_8$ scalar cosets are theories that admit
supersymmetric extensions, while the endpoints for all other cosets
are not the bosonic sectors of any supersymmetric theory.

\section*{Appendices}

\appendix

\section{Dimensional Reduction\label{sec:app}}

    Here, we give the general formulae for the toroidal reduction of
the Einstein-Hilbert Lagrangian in $D$ dimensions, coupled to a single
$p$-form field strength.  We consider a reduction on an $n$-torus.
The $D$-dimensional Lagrangian is
%%%%%
\be
{\cal L}_{\sst D} = R\, {*\oneone} -
\ft12 {*F_\p}\wedge F_\p .\label{Ddimlag}
\ee
%%%%%
The metric will be reduced using the standard Kaluza-Klein ansatz
which, in the notation of \cite{lpsol,cjlp1}, is
%%%%%
\be
ds_{\sst D}^2 = e^{\vec s\cdot\vec\phi}\, ds_{\sst D-n}^2 +
     \sum_{i=1}^n e^{2\vec\g_i\cdot\vec\phi}\, (h^i)^2\ ,
\ee
%%%%%
where
%%%%%
\be
h^i = dz^i + \cA^i_{\0 j}\, dz^j + \cA_\1^i= \td\g^i{}_j\, 
 (dz^j + \hA_\1^j)\ ,
\ee
%%%%%
where $\td\g^i{}_j = \delta^i_j + \cA^i_{\0 j}$.  We define also
$\g^i{}_j = (\td \g^{-1})^i{}_j$, as in \cite{cjlp1}.

   The constant vectors $\vec s$ and $\vec \g_i$ are given by
%%%%%%
\be
\vec s = (s_1,s_2,\ldots, s_n)\ ,\qquad
\vec\g_i = \ft12 \vec s -\ft12 \vec f_i\ ,
\ee
%%%%%
where
\be
s_i = \sqrt{\ft{2}{(\tD -1 -i)(\tD-2-i)}}\ ,\qquad
\vec f_i = \Big(\underbrace{0,0,\ldots, 0}_{i-1}, (\tD-1-i) s_i, s_{i+1},
s_{i+2}, \ldots, s_n\Big)\ ,
\ee
%%%%%
The potential $A_{\sst{(p-1)}}$ is reduced according to the standard
procedure
%%%%%
\be
A_{\sst{(p-1)}}\longrightarrow A_{\sst{(p-1)}} + A_{\sst{(p-2)i}}\, dz^i 
+ \ft12 A_{\sst{(p-1)ij}}\, dz^i\wedge dz^j \cdots \ .
\ee
%%%%%

     After reduction on the $n$-torus, the Lagrangian in $(D-n)$
dimensions is given by
%%%%%
\bea
{\cal L} &=& R\, {*\oneone} -\ft12 {*d\vec\phi}\wedge d\vec\phi 
-\ft12 e^{\vec a\cdot\vec\phi}\, {*F_\p}\wedge F_\p -\ft12 
\sum_i e^{\vec a_i\cdot\vec\phi}\, {*F_{\sst{(p-1)}i} }\wedge
F_{\sst{(p-1)}i}\nn\\
&& -\ft12 \sum_{i<j} e^{\vec a_{ij}\cdot\vec\phi}\, 
{*F_{\sst{(p-2)}ij} }\wedge F_{\sst{(p-2)ij}} -\cdots\nn\\
&& 
-\ft12 \sum_{i_1<i_2<\cdots <i_{p-1}} 
e^{\vec a_{i_1i_2\cdots i_{p-1}}\cdot\vec\phi}\, 
{*F_{\sst{(1)}i_1i_2\cdots i_{p-1}} }\wedge
 F_{\sst{(1)}i_1i_2\cdots i_{p-1}}\nn\\ 
&&-\ft12 \sum_i e^{\vec b_i\cdot\vec\phi}\, {*\cF_\2^i}\wedge \cF_\2^i
-\ft12 \sum_{i<j} e^{\vec b_{ij}\cdot\vec\phi}\, {*\cF^i_{\1 j}}\wedge
\cF^i_{\1 j} .\label{dmnlag}
\eea
%%%%%
The dilaton vectors are given by
%%%%%
\bea
&&\vec a = -(p-1)\, \vec s\ ,\qquad 
\vec a_i = \vec f_i-(p-1)\, \vec s\ ,\qquad
\vec a_{ij} = \vec f_i + \vec f_j -(p-1)\, \vec s\ ,\cdots\nn\\
&&\vec a_{i_1\cdots i_{p-1}} = \vec f_{i_1} + \vec f_{i_2} + \cdots 
 + \vec f_{i_{p-1}} -(p-1)\, \vec s \ ,\nn\\
&& \vec b_i = -\vec f_i\ ,\qquad \vec b_{ij} = -\vec f_i + \vec f_j 
\ .\label{dvec}
\eea
%%%%%
The field strengths are given by
%%%%%
\be
F_{\sst{(q)}i_1 i_2\cdots i_{p-q}} = \g^{j_1}{}_{i_1}\,
\g^{j_2}{}_{i_2} \cdots \g^{j_{p-q}}{}_{i_{p-q}}\, 
\hat F_{\sst{(q)}j_1 j_2\cdots j_{p-q}} \ ,\qquad
\cF^i = \td\g^i{}_j \hat{\cF}^j_\2\ ,
\ee
%%%%% 
with
%%%%%
\bea
\hat F_\p &=& dA_{\sst{(p-1)}} - dA_{\sst{(p-2)}i}\, \hA_\1^i +
  \ft12 dA_{\sst{(p-3)}ij}\, \hA_\1^i \, \hA_\1^j\nn\\
&& -\ft16 dA_{\sst{(p-4)}ijk}\, \hA_\1^i \, \hA_\1^j\, \hA_\1^k +
\cdots\ ,\nn\\
F_{\sst{(p-1)}i} &=&  dA_{\sst{(p-2)}} + dA_{\sst{(p-3)}ij}\, \hA_\1^j +
\ft12 dA_{\sst{(p-4)}ijk}\, \hA_\1^j \, \hA_\1^k\nn\\
&& +
\ft16 dA_{\sst{(p-5)}ijk\ell}\, \hA_\1^j \, \hA_\1^k\, \hA_\1^\ell 
+\cdots\ ,\nn\\
&&\cdots\cdots\nn\\
F_{\1 i_1\ldots i_{p-1}} &=& dA_{\0 i_1\ldots i_{p-1}}
\ .\label{fs1}\\
\hat{\cF}_\2^i &=& d\hat{\cA}_\1^i\ ,\qquad
\cF^i_{\1 j} = \gamma^k{}_j\, {\cal F}^i_{\1 k}\ .
\nn
\eea

    If we take the case where $D=n+3$, so that the reduction goes down
to three dimensions, the Lagrangian will be simply
%%%%%%%%%%%%
\bea 
{\cal L}_3 &=& R\, {*\oneone} -  \ft12{*d\vec\phi}\wedge d\vec\phi -
\ft12
\sum_i e^{\vec b_i\cdot \vec \phi}\, {*\cF}_\2^i\wedge \cF_\2^i\nn\\
&& -\ft12
\sum_{i<j} e^{\vec b_{ij} \cdot \vec\phi}\, {*\cF^i_{\1 j}}\wedge
\cF^i_{\1 j}
-\ft12 \sum_{i_1<\cdots<i_{p-2}} e^{\vec
a_{i_1\cdots i_{p-2}}\cdot \vec \phi}\, {*F}_{\2 i_1\cdots i_{p-2}}
\wedge F_{\2 i_1\cdots i_{p-2}}\nn\\
&& -\ft12 \sum_{i_1<\cdots<i_{p-1}}
e^{\vec a_{i_1\cdots i_{p-1}}\cdot \vec \phi}\, {*F}_{\1 i_1\cdots
i_{p-1}} \wedge F_{\2 i_1\cdots i_{p-1}}\ .  
\eea 
%%%%%%%%%
This is obtained from (\ref{dmnlag}) by dropping all field strengths
associated with forms of degree higher than 2.  We may then follow the
standard procedure for dualising the 1-form potentials  $\cA_\1^i$ and
$A_{\1 i_1\cdots i_{p-2}}$ to axionic scalars 
$\chi_i$ and $\psi^{j_1\cdots j_{p-2}}$, by introducing the axions as 
Lagrange multipliers
for the Bianchi identities for the 2-form field strengths (see, for
example, \cite{cjlp1}).   Upon doing so, we arrive at the purely
scalar three-dimensional Lagrangian
\bea
%%%%%%%%
{\cal L}_3 &=& R\, {*\oneone} -\ft12{*d\vec\phi}\wedge d\vec\phi
- \ft12
\sum_i e^{-\vec b_i\cdot \vec\phi}\, {*G}_{\1 i}\wedge G_{\1 i} \nn\\
&&-\ft12
\sum_{i<j} e^{\vec b_{ij} \cdot \vec\phi}\, {*\cF^i_{\1 j}}\wedge
\cF^i_{\1 j}
-\ft12 \sum_{i_1<\cdots<i_{p-2}} e^{-\vec
a_{i_1\cdots i_{p-2}}\cdot \vec \phi}\, {*G}_{\1}^{i_1\cdots i_{p-2}}
\wedge G_{\1}^{i_1\cdots i_{p-2}}\nn\\
&& -\ft12 \sum_{i_1<\cdots<i_{p-1}}
e^{\vec a_{i_1\cdots i_{p-1}}\cdot \vec \phi}\, {*F}_{\1 i_1\cdots
i_{p-1}} \wedge F_{\2 i_1\cdots i_{p-1}}\ ,\label{d3scallag}
\eea
%%%%%%%%%%
where
%%%%%%%%
\bea
F_{\1 i_1\cdots i_{p-1}} &=& \gamma^{j_1}{}_{i_1} 
\cdots \gamma^{j_{p-1}}{}_{i_{p-1}}\, dA_{\0 j_1 \cdots j_{p-1}}
\ ,\nn\\
G_{\1}^{i_1\cdots i_{p-2}} &=& \td\gamma^{i_1}{}_{j_1} 
\cdots \td\gamma^{i_{p-1}}{}_{j_{p-1}}\, d\psi_{\0}^{j_1 \cdots j_{p-1}}
\ ,\nn\\
G_{\1 i} &=& \gamma^j{}_i(d\chi_j - \sum_{k_1<\cdots< k_{p-2}}
A_{\0 k_1\cdots k_{p-2}j}\, d\psi^{k_1\cdots k_{p-2}})\ ,\nn\\
\cF^i_{\1 j} &=& \gamma^k{}_j\, d\cA^i_{\0 k}\ .\label{d3genkkmod}
\eea

\section{The $F_4$ oxidation endpoint and chiral $D=6$ supergravity}

    In general, the theories corresponding to the oxidation endpoints
of the various three-dimensional sigma-model Lagrangians that we have
considered in this paper are rather straightforward to analyse.  One
case is slightly more involved, namely the six-dimensional oxidation
endpoint of the $F_4$ scalar coset, discussed in section 3.4.  Here,
we present the details of the proof that it corresponds to the bosonic
sector of an $N=1$ chiral supergravity in $D=6$, with two
anti-self-dual tensor multiplets, and two abelian vector multiplets.

    From (\ref{d6sl2r}), we see that $G_\3$ satisfies $d{* G_\3} =
-\ft1{\sqrt2} d(\chi\, F_\3) - \ft12 dA_\1^+ \wedge dA_\1^-$, while
from (\ref{d6fields}), we see that the Bianchi identity for $dG_\3$
gives an identical expression.  This shows that we can indeed
consistently impose the self-duality constraint $G_\3={*G_\3}$ after
having obtained the equations of motion.

  The equation of motion for $F_\3$ is
%%%%% 
\be 
-d(e^{-\sqrt2\varphi}\, {*F_\3}) + \ft12 d(\chi^2\, F_\3) + \sqrt2\,
d(\chi\, G_\3) -\ft12 d(A_\1^+\, F_\2^+) + \ft1{2\sqrt2}\, d(\chi\,
A_\1^+\, F_\2^-) = 0\,,\label{feq}
\ee
%%%%%
which can be integrated to give
%%%%%
\be 
{*F_\3} = e^{\sqrt2\varphi}\, (dC_\2 + \sqrt2\, \chi\, G_\3 + \ft12
\chi^2\, F_\3 -\ft12 A_\1^+\, F_\2^+ + \ft1{2\sqrt2} \chi\, A_\1^+\,
F_\2^-)\,,
\ee
%%%%%
where we have introduced the third 2-form potential $C_\2$.  Thus, we
have
%%%%%
\bea
F_\3 &=& F_\3^-\,,\nn\\
G_\3 &=& F_\3^2 - \ft1{\sqrt2}\, \chi\, F_\3^-\,,\label{fields2}\\
{*F_\3} &=& e^{\sqrt2\varphi}\, (F_\3^+
 + \sqrt2\chi\, F_\3^2  - \ft12 \chi^2\, F_\3^- )\,,\nn
\eea
%%%%%
where we have defined
%%%%%
\be 
F_\3^-\equiv dA_\2 +\ft12 A_\1^-\, dA_\1^-\,,\quad
F_\3^2\equiv dB_\2-\ft12 A_\1^+\, dA_\1^-\,,\quad
F_\3^+\equiv dC_\2-\ft12 A_\1^+\, dA_\1^+\,.\label{f3defs}
\ee
%%%%%

  We wish to make a comparison with the chiral $N=1$ six-dimensional
supergravity described in \cite{sagnotti}.  In \cite{sagnotti}, there
is  1 self-dual 3-form $H_\3$, and $n$ anti-self-dual 3-forms $K^m_\3$.  We
have the opposite convention, so we have 1 anti-self-dual
$H_\3$, and, in our case, $n=2$ self-dual 3-forms $K^m_\3$, with
$m=1,2$.  We may relate these fields
and our fields $G_\3$ and $F_\3$, as follows:
%%%%%
\bea
H_\3 &=& \ft12 e^{-\fft1{\sqrt2}\varphi}\, (F_\3-{*F_\3})\,,\nn\\
K^1_\3 &=& \ft12 e^{-\fft1{\sqrt2}\varphi}\,(F_\3 + {*F_\3})
\,,\label{hkdefs}\\
K_\3^2 &=& G_\3\,.\nn
\eea
%%%%%
We can now express these in terms of our three field strengths
defined in (\ref{f3defs}), by making use of the results in (\ref{fields2}).
In fact it is advantageous at this point to
define
%%%%%
\be
F_\3^- \equiv F_\3^1 + F_\3^0\,,\qquad F_\3^+\equiv
F_\3^1-F_\3^0\,,\qquad  F_\3^2\,.\label{rotate}
\ee
%%%%%
The fields $F_\3^0$, $F_\3^1$ and $F_\3^2$ will now correspond
to the $F_\3^r$ ($r=0,1,2$) fields of \cite{sagnotti}.  Thus we have
%%%%%
\bea
H_\3&=& \cosh\fft{\varphi}{\sqrt2}\, F_\3^0
-\sinh\fft{\varphi}{\sqrt2}\, F_\3^1 - \ft1{\sqrt2}\chi\,
e^{\fft1{\sqrt2}\, \varphi}\, F_\3^2 +\ft14 e^{\fft1{\sqrt2}\varphi}\,
\chi^2\, (F_\3^0 + F_\3^1) \,,\nn\\
K_\3^1 &=& \cosh\fft{\varphi}{\sqrt2}\, F_\3^1
-\sinh\fft{\varphi}{\sqrt2}\, F_\3^0 + \ft1{\sqrt2}\chi\,
e^{\fft1{\sqrt2}\, \varphi}\, F_\3^2 -\ft14 e^{\fft1{\sqrt2}\varphi}\,
\chi^2\, (F_\3^0 + F_\3^1) \,,\nn\\
K_\3^2 &=&  F_\3^0 - \ft1{\sqrt2}\, \chi\, (F_\3^0 + F_\3^1)\,.\label{ours}
\eea
%%%%%

   This can be directly compared with equations (3.5) in
\cite{sagnotti}, namely
%%%%%
\be
H_\3 = v_r\, F_\3^r\,,\qquad K^m_\3 = x^m{}_r\, F_\3^r\,,
\ee
%%%%%
allowing us to read off the quantities $v_r$ and $x^m{}_r$.  We find
that the matrix $V$ defined in \cite{sagnotti}, \ie 
%%%%
\be
V = \pmatrix{ v_0&v_1&v_2\cr
              x^1{}_0& x^1{}_1 & x^1{}_2\cr
              x^2{}_0 & x^2{}_1 & x^2{}_2 }\,,
\ee
%%%%%
is given by
%%%%% 
\be 
V = \pmatrix{\cosh\fft{\varphi}{\sqrt2} +
\ft14 e^{\fft1{\sqrt2}\varphi}\, \chi^2 & -\sinh\fft{\varphi}{\sqrt2}
+\ft14 e^{\fft1{\sqrt2}\varphi}\, \chi^2 & -\ft1{\sqrt2}
e^{\fft1{\sqrt2}\varphi}\, \chi \cr
  -\sinh\fft{\varphi}{\sqrt2} -
\ft14 e^{\fft1{\sqrt2}\varphi}\, \chi^2 &
\cosh\fft{\varphi}{\sqrt2} -
\ft14 e^{\fft1{\sqrt2}\varphi}\, \chi^2
& \ft1{\sqrt2} e^{\fft1{\sqrt2}\varphi}\, \chi\cr
-\ft1{\sqrt2}\chi & -\ft1{\sqrt2}\, \chi & 1}\,.\label{vres}
\ee
%%%%%
    
    It is easily verified that $V$ is an $SO(1,2)$ matrix, satisfying
$V^T\, \eta\, V = \eta$, where $\eta={\rm diag}\, (1,-1,-1)$.  It can
also be seen that
%%%%%
\be 
dV\, V^{-1} = \fft1{\sqrt2}\, \pmatrix{0& -d\phi & - e^{\fft1{\sqrt2}
\varphi} \, d\chi \cr
   -d\phi & 0 &  e^{\fft1{\sqrt2} \varphi}\, d\chi \cr
- e^{\fft1{\sqrt2}\varphi} \, d\chi & - e^{\fft1{\sqrt2}
\varphi} \, d\chi & 0}\,.
\ee
%%%%%

   In \cite{sagnotti}, the field strengths $F_\3^r$ are defined to
be  
%%%%%
\be
F_\3^r = dA_\2^r + c^r_z\, \omega^z\,,\label{fomdef}
\ee 
%%%%%
summed over the various factors in the Yang-Mills gauge group,
labelled by $z$, where $\omega^z$ are the Chern-Simons forms for each
factor and $c^r_z$ are constants.  In our case, the situation will be
slightly different: we have two $U(1)$ factors, with 1-form potentials
that we may call $A_\1^+$ and $A_\1^-$; they form a spinor of
$SO(1,2)$. We shall need a third vector combining the previous two,
namely
%%%%%
\be
A_\1 = A_\1^- - A_\1^+\,.\label{trunc}
\ee
%%%%%
Actually the $SO(1,2)$ covariance will be preserved by choosing for the 
coefficients $c^r_z$ the Clebsch-Gordans of this group.\footnote{In fact
it is natural to extend the class of theories discussed in
\cite{sagnotti}, at least for abelian vector fields, by considering
Chern-Simons corrections to the 3-forms with the more general
structure $F_\3^r = dA_\2^r + c^r_{\a\b} \, A^\a_\1\, dA_\1^\b$.  The
constants $c^r_{\a\b}$ in our case are then precisely the
Clebsch-Gordan coefficients relating one spin-1 to the product of two 
spin-$\ft12$ $SL(2,\R)$ representations.  We believe that the proof of
supersymmetry for the ``diagonal'' cases $c^r_z \sim c^r_{\a\a}$
discussed in \cite{sagnotti} extends straightforwardly to these more
general structures.}  The
 corresponding three Chern-Simons forms are
%%%%%
\be
\omega^+ = A_\1^+\, dA_\1^+\,,\qquad \omega^- = A_\1^-\,
dA_\1^-\,,\qquad
\omega^3 = A_\1\, dA_\1\,.\label{omegadef}
\ee 
%%%%%
(In each case, one is free to add any total derivative to $\omega$,
since the only important point is that $d\omega = F_\2\wedge F_\2$ in
each factor.)

Note that $A_\1^+$ and $A_\1^-$ are precisely the two $U(1)$ gauge
potentials appearing in our Lagrangian (\ref{d6sl2r}).  From
(\ref{f3defs}) and (\ref{rotate}), we have
%%%%%
\bea
F_\3^0 &=& \ft12 d(A_\2-C_\2) + \ft14 A_\1^-\, dA_\1^- + \ft14 A_\1^+
\, dA_\1^+\,,\nn\\
F_\3^1 &=&  \ft12 d(A_\2+C_\2) + \ft14 A_\1^-\, dA_\1^- - \ft14 A_\1^+
\, dA_\1^+\,,\nn\\
F_\3^2 &=& d(B_\2 + \ft14 A_\1^+ A_\1^-) - \ft14 A_\1^+\, dA_\1^- - \ft14 A_\1^-\, dA_\1^+\ ,.
\eea
%%%%%
Comparing with (\ref{fomdef}) and (\ref{omegadef}), bearing in mind
(\ref{trunc}), we can read off the values of the constants $c^r_z$ for
each $r=(0,1,2)$ and each $z=(+,-,3)$, leading to
%%%%%
\bea
&&c^0_+ =\ft14\,,\quad c^0_- = \ft14\,,\qquad c^0_3 =0\,,\nn\\
&&c^1_+ =-\ft14\,,\quad c^1_- = \ft14\,,\qquad c^1_3 =0\,,\label{cres}\\
&&c^2_+ =-\ft14\,,\quad c^2_- = -\ft14\,,\qquad c^2_3 = \ft14\,,\nn
\eea
%%%%%
For the purposes of comparison with \cite{sagnotti}, we have used the
diagonal constants $c^r_z$ in eqn (\ref{cres}).

   This completes our comparison of the theory described by
(\ref{d6sl2r}) with the family of chiral $D=6$ supergravities
described in \cite{sagnotti}).  It corresponds, as we have seen, to the
situation where the supergravity multiplet, with its anti-self-dual
tensor multiplet, is supplemented by two self-dual tensor matter
multiplets, and two abelian vector matter multiplets.  The structure
of the Chern-Simons terms is a slight modification of that given in
\cite{sagnotti}.

\section*{Acknowledgments}

    C.N.P. is grateful to the \'Ecole Normale for hospitality during
the course of this work.

\end{document}